\documentclass[12pt]{iopart}
\usepackage{epsfig}

\begin{document}

\title{A multi-detector array for high energy nuclear
\mbox{$\bf e^+\bf e^-$ pair spectrosocopy}} 

\author{K E Stiebing\S, F W N de Boer\P, O Fr\"{o}hlich\S, H Bokemeyer\ddag, \newline
K A M\"{u}ller\S, K Bethge\S \,and J van Klinken$\parallel$}
\address{\S\  Institut f\"{u}r Kernphysik der
Johann Wolfgang Goethe-Universit\"{a}t (IKF),\\
D-60486 Frankfurt/Main, Germany}
\address {\P\  Nationaal Instituut voor Kernfysica en Hoge-Energie Fysica
(NIKHEF),\\ 1009 DB Amsterdam, the Netherlands}
\address{\ddag\   Gesellschaft f\"{u}r Schwerionenforschung (GSI), D-64291
Darmstadt, Germany}
\address{$\parallel$\ Kernfysisch Versneller Instituut (KVI),
9747 AA Groningen, The Netherlands}
\eads{\mailto{\tt stiebing@ikf.uni-frankfurt.de}; or: \mailto{\tt
fokke@nikhef.nl}}

\begin{abstract}
A multi-detector array has been constructed for the simultaneous measurement
of energy- and angular correlation of electron-positron pairs produced
in internal pair conversion (IPC) of nuclear transitions up to 18 MeV.
The response functions of the individual detectors have been measured
with mono-energetic beams of electrons. Experimental results obtained with
1.6 MeV protons on targets containing $^{11}$B and $^{19}$F show clear
IPC over a wide angular range. A comparison with GEANT simulations demonstrates that angular correlations of $e^+e^-$ pairs of transitions in the energy 
range between 6 and 18 MeV can be determined with sufficient resolution and 
efficiency to search for deviations from IPC due to the creation and subsequent decay into $e^+e^-$ of a hypothetical short-lived neutral boson.
\end{abstract}
\submitto{Journal of Physics G} 

\newpage

\section{Introduction}
\label{sec:intro}
Spectroscopy of internal pair conversion (IPC), has a long
tradition \cite{wilson-siegbahn}. It has been an essential tool
to identify nuclear transitions with energies in the few MeV region.
This paper describes an experimental arrangement which has been used in
a search for elusive neutral bosons. In the investigations with this
equipment we obtained positive signals for deviations from IPC which may be of basic interest.     

The IPC-branching ratio relative to the competing process of
${\gamma}$-ray emission and the angular correlation
of the pairs depend on the multipolarity of the transition and 
measurements of IPC have enabled an accurate determination of
 energy, spin and parity of many nuclear levels.
Magnetic spectrometers have been
used to identify coincident $e^+e^-$ pairs
in order to measure the multipolarity and, with high precision,
the energy of the corresponding transition.
However, the excellent energy resolution of magnetic spectrometers can only
be achieved at the expense of a small acceptance and hence long counting
times were needed for such measurements.

For high-statistics measurements of the angular correlation of
$e^+e^-$ pairs, primarily detector telescopes with large solid angles
were used. Each telescope consists of two detectors, usually a
combination of a thin ($\Delta$E) and a thick (E) detector operated in
coincidence to suppress singles ${\gamma}$-ray events
(e.g. from Compton processes). The angular correlation was then measured
stepwise with two telescopes placed at different angles or,
preferably, simultaneously by a multi-telescope assembly.

Following the early experience obtained in the fifties and sixties,
several new large acceptance $e^+e^-$ pair spectrometers have
been constructed in the nineties in particular at KVI
Groningen \cite{budaThesis,budaNIM,buda-nucl-phys,budaphysrev,budaphysrevlett}
and Stony Brook \cite{montoya,schadmand1,schadmand2}.

Some years ago, a search for neutral bosons \cite{fokke0}, possibly emitted in
nuclear transitions and decaying into $e^+e^-$ pairs, has been
started offering a new challenge to old techniques. Within
the constraints of energy and spin-parity conservation,
such a neutral boson 
would compete with ${\gamma}$-ray and IPC decay in certain
nuclear transitions. The signature of such a boson would be its
two-body decay into an electron-positron pair. 

IPC occurs when a nucleus emits an $e^+e^-$ pair instead of a ${\gamma}$-ray 
quantum. Together with the recoil of the emitting nucleus this forms
a planar three-body decay with usually a small $e^+e^-$ opening angle, 
E0 transitions forming a noticeable exception to this rule. 
However, in the conceivable case when the nucleus emits a neutral boson,
instead of IPC, this happens by two-body decay. When the boson subsequently 
decays into an $e^+e^-$ pair, again a two-body decay occurs, resulting in 
back-to-back emission of the $e^+e^-$ pairs in the centre of mass system. 
In the laboratory system the momentum of the decaying boson makes 
this angle smaller than 180$^{o}$ but leaves the  
${\Theta}$ distribution totally different from IPC: the ${\Theta}$ distribution becomes truly instrumental in searches for such elusive neutral particles.  


Various experimental studies have provided constraints on the
possible existence of light, elusive neutral bosons
\cite{gunion,rev}. Short-lived neutral bosons have been searched
for in beam dump experiments \cite{riordan,davier,bross} for the
mass region between the $e^+e^-$ threshold at 1.022 MeV/c$^2$
and the mass of the muon (105.7 MeV/c$^{2}$). It is interesting to
note that these searches still allow a mass-lifetime window for
masses between 5 and 100 MeV/c$^{2}$ at lifetimes shorter than
10$^{-13}$\, s.

A decade ago our interest in the possible existence of a hitherto
undiscovered neutral boson was also triggered by results from
emulsion studies \cite{elnadibadawy,deboervandantzig} 
of relativistic heavy ion reactions, in which 
a distinct cluster of $e^+e^-$ pairs was observed 
at short distance from the interaction vertex. 
Interpreted in terms of the emission and
subsequent decay of a light $X$ boson, the relevant events show an
average invariant mass $m_X$ of ${\sim}$ 9 MeV/c$^{2}$ and lifetime
${\tau}_X$ $\sim 10^{-15}$ s.

The possibility to investigate the open mass-lifetime window provided
the motivation \cite{froehlich} to build an apparatus to perform
high-statistics IPC measurements of E1 and M1 transitions and to
search for anomalies in the angular correlations of $e^{+}e^{-}$
pairs with invariant mass between 5 and 15 MeV/c$^{2}$
 and a lifetime shorter than {\mbox{$10^{-10}$ s.} 

Here, we report on the design and construction of the
apparatus at the Institut f\"ur Kernphysik, Frankfurt (IKF)
and describe some experiences with the equipment during a number of
experiments \cite{fokke0,fokke1,fokke2,fokke3} which demonstrate its 
performance.

\section{General considerations}
\label{sec:general}

In order to investigate deviations from normal internal
pair-conversion, a thorough understanding of the apparatus and its
properties has to be achieved and demonstrated. Therefore,
extended simulations and calibration procedures were performed during several
stages of the construction of the spectrometer. 
The simulations were performed with the GEANT code from CERN\cite{geant}. 

Besides the IPC process, also the background of external pair creation (EPC) and
multiple lepton scattering must be considered. For this purpose
several event generators were developed and applied to the code. 
Below we give a survey of the most important generators
used in this implementation. 
A more detailed description is given in the PhD thesis of Fr\"{o}hlich
\cite{froehlich}.

\subsection{Internal pair conversion (IPC)}

For E1 and  M1 transitions the formulation by Goldring
\cite{goldring} for the triple differential pair conversion
coefficient is used. In the Born approximation for E1 and M1
multipolarities the differential transition probabilities are:

\begin{eqnarray}                    
\frac{d{\Gamma}}{d{\Omega}_{+}d{\Omega}_{-}dE_{+}}&&({\Theta},E_{+})
= \ \ \ \ \ \ \ \ \ \ \ \ \ \ \ \ \ \ \ \ \ \ \ \ \ \ \ \ \ \ \ \
\ \ \ \nonumber\\
&&\frac{{\alpha}}{16{\pi}^{3}k^{3}}\,p_{+}p_{-}
\left\{\frac{4k^{2}}{(k^2 - q^2)^2} + 2 \frac{p^2_+ + p^2_-}
 {k^2 - q^2}  \pm 1 \right\} 
\label{eq:p57-3_4a}
\end{eqnarray}
with the $\pm$ sign equal to $-$ for M1 and $+$ for E1.
The parameter $k\, {\equiv}\, E_+ + E_-$ is the pair energy in units of 
$m_{e}c^2$, 
$\vec{q}$\, ${\equiv}$\, $\vec{p}_{+} + \vec{p}_{-}$ is the total momentum in units 
of $m_{e}c$\,, and $\alpha$ is the fine structure constant.

The square of the invariant mass of the electron positron pair,  
${\mu}^2 = k^2 - q^2$, depends on the opening angle $\Theta$
according to 
\begin{equation}                                               
{\mu}^{2} =  2\,(E_+E_- +1)\,(1 - u\cos{\Theta})
\label{eq:p58-nonuma}
\end{equation}
where $u\,{\equiv}\,p_{+}p_{-}/(E_{+}E_{-} + 1)$
is close to unity
for sufficiently high pair energies.
%
%
The $\Theta$ dependence of the differential pair conversion coefficient 
in Eq.\,\ref{eq:p57-3_4a} is mainly determined by the second term and is similar 
for E1 and M1 transitions over a wide range of correlation angles. 
It has the approximate shape of $(1 - \cos\,{\Theta})^{-1}$. 

For E0 transitions the differential production cross section
for pair conversion can be written in first order as
\cite{soff,hofmann1,hofmann2,hofmann3}:

\begin{equation}                                       
\frac{dP}{dE_+d\cos{\Theta}} =
\frac{1}{2}\,\frac{dP}{dE_+}\,[1+\epsilon\cos{\Theta}]
\label{eq:p68-3_14}
\end{equation}
Here ${\epsilon}$ is an anisotropy factor, given in Born
approximation for light nuclei by:

\begin{equation}                                       
{\epsilon} = \frac{p_+p_-}{E_+E_- -1}. 
\label{eq:p68-3_15a}
\end{equation}
Equations 2 and 3 are used as input for a two dimensional random
generator, provided by the HBOOK library (CERN)\,\cite{geant},
to generate the emission angle and the energy of the positron in an IPC event.

\subsection{External pair conversion (EPC)}
\label{sec:EPC}

Except for monopole transitions, all excited nuclear levels decay
predominantly by ${\gamma}$-ray emission. Therefore special
attention has to be paid to the background created by external
pair conversion (EPC) in the
spectrometer\,\cite{borsellino,olsen,hart,hubbell}. For E$_{\gamma}$ between 10
and 20\,{MeV} and for $Z$ values smaller than 20 the cross
section for EPC on a nucleus can be approximated by:

\begin{equation}                                 
{\sigma}_{EPC}\,{\approx}\, 1.531\, Z^{2} \ln {\frac{E_{\gamma}}
{2.67 \, {\rm  MeV}}}  \,\,\,\, {\rm mb}.   
\end{equation}

The cross-section for EPC in the field of the atomic electrons is
a factor $C/Z$ smaller, with $C \approx\,$1.13 at
17\,{MeV} transition energy. It amounts to about 15\% for $^{12}$C
and therefore should be added to the contribution by nuclear EPC.

The standard implementation of EPC in GEANT consists of two steps:
i) selection of the first lepton's energy by an
acceptance-rejection method (von Neumann)\,\cite{rev} within the
constraints of the Bethe-Heitler theory, ii) selection of the
opening angle between the leptons by using an approximation to the 
theory by Tsai\,\cite{tsai}. 
The momentum vectors of the leptons and the recoiling nucleus are 
then given by momentum conservation.

The comparison of the energy- and angle distributions of
this implementation to results obtained by the theory of
Olsen\,\cite{olsen}, which is more appropriate for the transition
energies considered here, shows significant differences. The
standard implementation results in a shift of the maximum of the
energy distribution by a factor $\sim${2} and in a more
moderate slope of the opening angle distribution. This 
leads to an overestimation \cite{froehlich} of larger opening angles by a factor of up to $\sim${10}. Therefore, the standard
GEANT routines were modified to include an option to create lepton
pairs according to the theory of Olsen \cite{olsen}.

The essential difference between EPC and IPC stems from the fact
that EPC depends on the geometry of the setup. As a result the
simulation of EPC requires at least two orders of magnitude more
computing time than for IPC. 
To keep the computing time within reasonable limits, the probability
for EPC was increased by a factor 1400 in our simulations. 

The angular correlation for EPC is strongly
peaked around ${\Theta} = 0^{o}$ and is well described by 
\mbox{$N(\Theta) = N_0$ exp\,$(-\Theta/\tilde{\Theta})$.} Using the option for 
the Olsen theory results in
$\tilde{\Theta} = 8.4^{o}$ degrees, whereas the standard GEANT option 
yields $\tilde{\Theta} = 9.0^{o}$.

\subsection{The two-body decay of a boson}
\label{sec:two-body}

For our boson search it is crucial that the disintegration of the boson into an $e^+e^-$ pair is a 
two body decay. In the centre of mass system (CM) of the  
boson the $e^+$ and $e^{-}$ are emitted under 180$^{o}$. In the
laboratory system the boson moves with ${\beta} {\leq} 0.9$,
resulting in opening angles which can be much smaller than 180$^{o}$.
 
In the simulations, the emission of the $X$ boson as well as its decay 
into a lepton pair were assumed to be isotropic in the respective CM systems.
The invariant mass $m_{X}$ of the boson, the 
positron and electron momenta $\vec{p}_+$ and $\vec{p}_-$ and their energies 
$E_+$ and $E_-$ are related by

\begin{equation}
m_X^2 = 2m_e^2 + 2E_+E_- - 2|\vec{p}_{+}||\vec{p}_{-}|\cos{\Theta}, 
\end{equation}
where ${\Theta}$ is the angle between the electron and positron momenta.

\section{Apparatus}
\subsection{Detector system}

The IKF spectrometer consists of six detector telescopes. They are
mounted at fixed angles allowing the simultaneous observation of
15 correlation angles. Each telescope consists of a ${\Delta}$E-detector
($2.2{\times}2.2$ cm$^{2}$; thickness: 0.1 cm), and
an E-detector  ($3.0{\times}3.0$ cm$^{2}$;
thickness: 7.0 cm). The
telescopes are positioned on a circular rail with the beam axis as
center. In order to allow for a differential study of the energy split
of the lepton pairs, this basic array has been extended by two ``larger"
telescopes for collecting better statistics (${\Delta}$E-detectors:
$3.8{\times}4.0$ cm$^{2}$ and thickness: 0.3 cm; E-detectors:
$8.0{\times}6.0$ cm$^{2}$ and thickness 7.0 cm). In the energy region
of interest (5 to 20 MeV) the energy loss of the electrons is
roughly constant and amounts to ${\sim}$ 200 keV for the 1 mm and 600 keV
for the 3 mm ${\Delta}$E-detectors \cite{pages}.
The two larger
 detectors can be
mounted to cover a selected range of correlation angles, e.g. where deviations 
from the decay pattern of conventional IPC were observed. 
Their implementation into the basic array raised the number of
correlation angles to 28.

By displacing the target along the beam axis changes in the relative angles 
between the detectors remained to some extent still possible. 
The distances from the ${\Delta}$E-detectors to the target are about
11 and 14 cm for the small and the large detectors respectively.

\begin{figure}[htb]  
\begin{center}
\epsfig{file=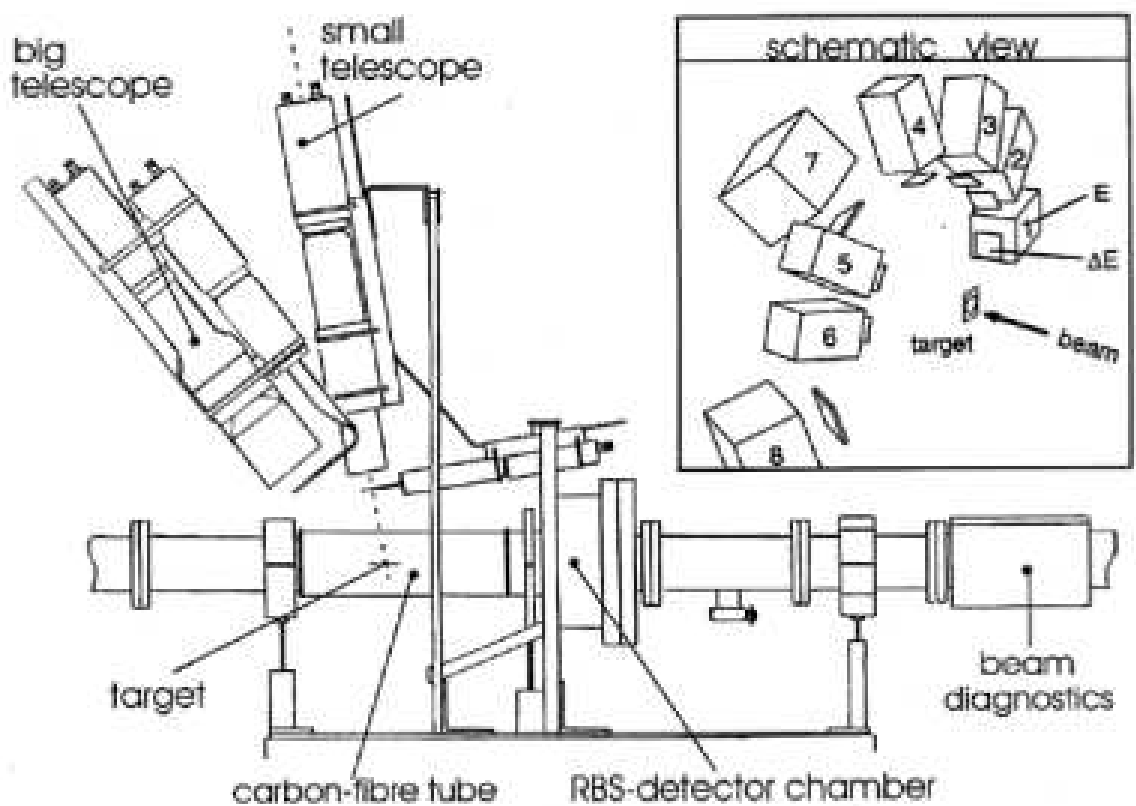,width=\textwidth}
\end{center}
\caption[sketch]{
 Side view of the experimental
 setup consisting of six small and two large ${\Delta}E-E$ telescopes,
 the beam tube, a Rutherford backscattering (RBS) device
 to monitor the target condition and a carbon-fibre tube
 as target chamber. For clarity, only one large and one small
 telescope are depicted. The insert shows a perspective view of all telescopes.}
\label{fig:sketch}
\end{figure}

\subsection{Beam and Target}

In Fig.\,\ref{fig:sketch} a side view of the experimental setup is sketched.
The beam, coming from the right hand side, first passes a beam
diagnosis chamber.
To minimize the amount of material around the target, a 24{\,}cm long
electrically conducting carbon fibre tube 
with a radius of 3.5{\,}cm and a wall thickness of only 0.8{\,}mm is used.
The target, positioned perpendicular to the beam, was held by a single rod of 
3 mm thickness at the side opposite to the spectrometer system. Aluminum target frames of a thickness of only 
0.3{\,}mm had outer dimensions of 20$\times$25{\,}mm$^2$ and an
inner diameter of 15{\,}mm. Target foils were thin: typically less than 2 mg/cm$^{2}$. Their thickness was either adapted to the resonance width of the reaction under investigation  or to the 
demand of a sufficient true-to-random ratio of coincidences. The beam is stopped in a tantalum cup at 150 cm behind the target.
In order to achieve approximately the same average path length inside the target for all detected leptons,
the detectors were installed at slightly inclined angles (by 25$^{o}$ for the 
small telescopes) relative to the target plane. In this way, interference by the
target frame is avoided with only a moderate reduction of the
correlation-angular range.

\begin{table}[hbt]
\begin{center}
\begin{minipage}{14cm}
\caption[Experimental]{
 Position, solid angles and orientation of the detector telescopes relative to 
the target. The accuracy in the distances and angles is estimated
to be about one millimeter and two degrees.} 
\label{tab:detang}
\vspace{2.mm}
\begin{tabular}{cccccc}
\hline
 Detector & Distance  & Solid angle & Polar & Azimuthal \\
  & && angle & angle     \\
 \hline
& [cm] &  [10$^{-2}$\,sr]&&\\
 \hline
 1 & 11.50 &3.57 & 174.6$^{o}$ & 63.4$^{o}$ \\
 2 & 12.00 &3.28 & 150.5$^{o}$ & 64.6$^{o}$ \\
 3 & 11.50 &3.56 & 127.8$^{o}$ & 63.4$^{o}$ \\
 4 & 11.65 &3.48 & 86.3$^{o}$ & 63.8$^{o}$  \\
 5 & 10.70 &4.12 & 26.2$^{o}$ & 66.3$^{o}$  \\
 6 & 10.93 &3.87 & 0.34$^{o}$ & 64.3$^{o}$  \\
 7 & 13.02 &9.57 & 49.7$^{o}$ & 41.7$^{o}$  \\
 8 & 15.20 &8.76 & -34.2$^{o}$ & 64.6$^{o}$ \\
\hline
\end{tabular}
\end{minipage}
\end{center}
\vspace{2.mm}
\end{table}

In Table\,\ref{tab:detang} the azimuthal and polar angles of the
detectors are listed together with the distances to the target. 
In Table\,\ref{tab:relang} the relative central correlation angles
are listed for all 28 detector combinations. The rather high redundancy in
correlation angles provides an independent way to examine the
detector efficiencies (see section \ref{sec:coeldaq}).

\begin{table}[htb]
\begin{center}
\caption[Experimental2]{
 Correlation angles defined by the telescope positions as given
 in Table 1.} 
 \label{tab:relang} \vspace{2.mm}
\begin{minipage}[t]{3cm}
\begin{tabular}{cc}
\hline
 Relative  & Det.\\
 angle  & comb.\\
\hline
 20.6$^{o}$  & 2-3\\
 21.8$^{o}$  & 1-2\\
 23.8$^{o}$  & 5-6\\
 31.3$^{o}$  & 5-7\\
 31.3$^{o}$  & 6-8\\
 36.4$^{o}$  & 4-7\\
 37.2$^{o}$  & 3-4\\
\hline
\end{tabular}
\end{minipage}
\hfill
\begin{minipage}[t]{3cm}
\begin{tabular}{cc}
\hline
 Relative  & Det.\\
 angle  & comb.\\
\hline
 42.0$^{o}$ & 1-3\\
 44.9$^{o}$ & 6-7\\
 54.6$^{o}$ & 4-5\\
 54.9$^{o}$ & 5-8\\
 57.6$^{o}$ & 2-4\\
 63.6$^{o}$ & 3-7\\
 68.0$^{o}$ & 7-8\\
\hline
\end{tabular}
\end{minipage}
\hfill
\begin{minipage}[t]{3cm}
\begin{tabular}{cc}
\hline
 Relative  & Det.\\
 angle  & comb.\\
\hline
 76.4$^{o}$ & 4-6\\
 77.9$^{o}$ & 1-4\\
 79.0$^{o}$ & 2-7\\
 90.0$^{o}$ & 3-5\\
 91.6$^{o}$ & 1-7\\
 103.8$^{o}$ & 4-8\\
 108.4$^{o}$ & 3-6\\
\hline
\end{tabular}
\end{minipage}
\hfill
\begin{minipage}[t]{3cm}
\begin{tabular}{cc}
\hline
 Relative  & Det.\\
 angle  & comb.\\
\hline
 108.3$^{o}$ & 2-5\\
 122.8$^{o}$ & 1-5\\
 122.9$^{o}$ & 2-6\\
 122.4$^{o}$ & 1-8\\
 126.7$^{o}$ & 3-8\\
 129.3$^{o}$ & 1-6\\
 130.6$^{o}$ & 2-8\\
\hline
\end{tabular}
\end{minipage}
\end{center}
\vspace{2mm}
\end{table}

As the lepton ranges in the E-detectors vary strongly with
energy, a Lucite light guide has been inserted between scintillator and
photomultiplier in order to make the detected pulse height
independent of the position of the scintillation light emission. 
These light guides had a length of 10 \,{cm} for the small and 5 \,{cm} for the
large telescopes, respectively. The dependence of the E-detector's
signals on the position of the photon emission was checked in
a series of measurements with a collimated $^{207}$Bi source and a
magnetically analyzed $^{90}$Sr source. The measured pulse heights
for any position on the long side (detector depth) of a scintillator block were constant within 1\% for
all E-detectors of the small telescopes \cite{froehlich}. Also the 
${\Delta}$E-detectors were coupled via light guides to their photomultipliers.

During prolonged beam-exposure the target may deteriorate by sputtering 
or by contamination with decomposing or condensing vacuum remnants. 
For on-line monitoring of the status of the target the setup was
complemented with a surface barrier detector for Rutherford
backscattering spectroscopy (RBS) mounted at 175$^{o}$ scattering angle. 
By comparing the continuously monitored RBS spectra during the measuring 
periods, any change in the homogeneity of the target
can immediately be identified by a deterioration of the peak to
background ratio and/or a broadening of the RBS
structures \cite{froehlich}. In contrast to the natural boron,
the LiO$_2$ targets for the $^7$Li(p$,e^+e^-)^8$Be reaction
had to be exchanged once every two hours. 
During this time the Li-backscatter signal typically dropped below 50\%.

\begin{figure}[htb] 
\leavevmode
\begin{center}
\epsfig{file=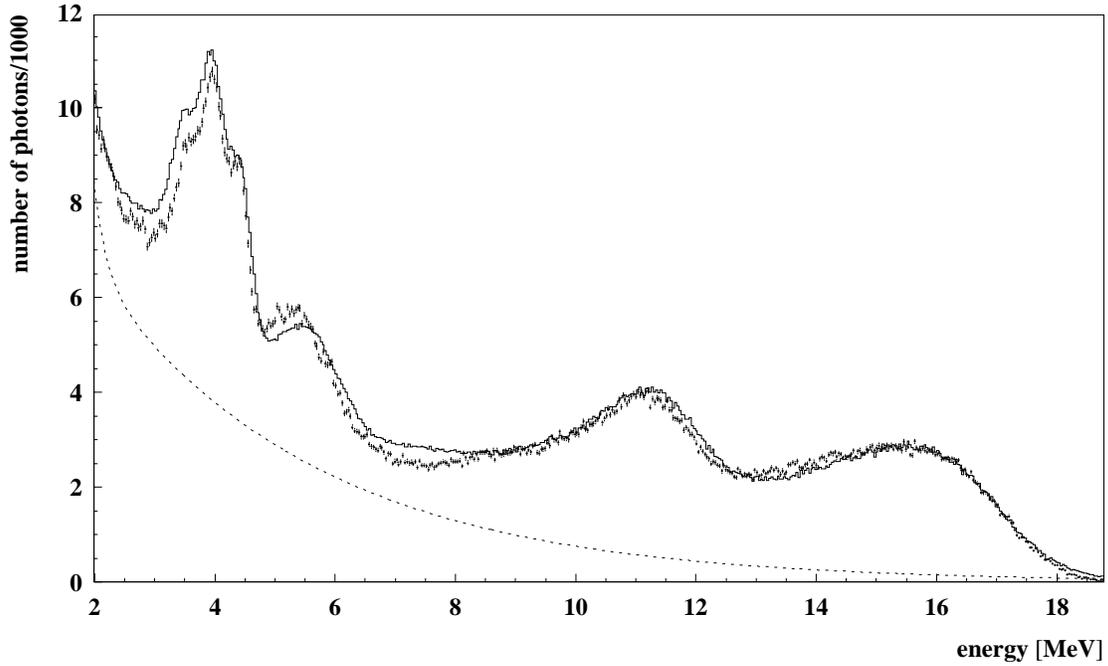,width=\textwidth}
\end{center}
\caption[]{
 Gamma-ray spectrum from natural boron bombarded with 1.6 MeV protons
 measured with the NaI detector (symbols).
 The histogram represents a GEANT-simulation. 
The "empty target" background, approximated by an exponential
function(dashed line) has been added to the GEANT simulation.}
\label{fig:ngm}
\end{figure}

In order to provide a means to compare the measured conversion
spectra to the corresponding $\gamma$-spectra, a 3"$\times$3"
NaI(Tl) detector (not shown in Fig.\,\ref{fig:sketch}) was located
at 53.3 cm vertically from the beam axis covering a solid angle of
16.3\, msr.
The detector response function has been measured
up to energies of 2.7 MeV and has been extrapolated to higher
energies assuming an energy dependent Gaussian line shape
plus tails (see Fig.\,2).
Its total detection efficiency in the energy range considered here
has been obtained by GEANT\, simulations.
Fig.\,\ref{fig:ngm} demonstrates that the GEANT-simulation
adequately reproduces the ${\gamma}$-ray spectrum following the
bombardment of a natural boron target with 1.6 MeV protons. For
cases where a better resolution of the $\gamma$-ray detection is
necessary, the set up was complemented by a Ge(Li) detector.

\subsection{Line shapes of the E-detectors}
\label{sec:lineshape}

The response of the detectors to monochromatic leptons was
investigated at the linear electron accelerator (LINAC) of the
Strahlenzentrum Giessen. The electron beam was collimated to a
spot of (1.0$\times$1.0) cm$^{2}$, and had an energy spread of less
than 5\% FWHM.
The beam entered the telescopes in normal condition, i.e. 
through the ${\Delta}$E-detectors into the E-detectors, along the 
telescope axis.
As a typical example Fig.\,\ref{fig:linshap} shows the measured
response of an E-detector to electrons 
at six different energies in the range of 6 to 18 MeV. The histograms in 
the left column of Fig.\,\ref{fig:linshap} represent GEANT-simulations of 
this  ``Giessen setup" assuming an energy-dependent Gaussian energy resolution
function. 

\begin{figure}[htb] 
\leavevmode
\begin{center}
\epsfig{file=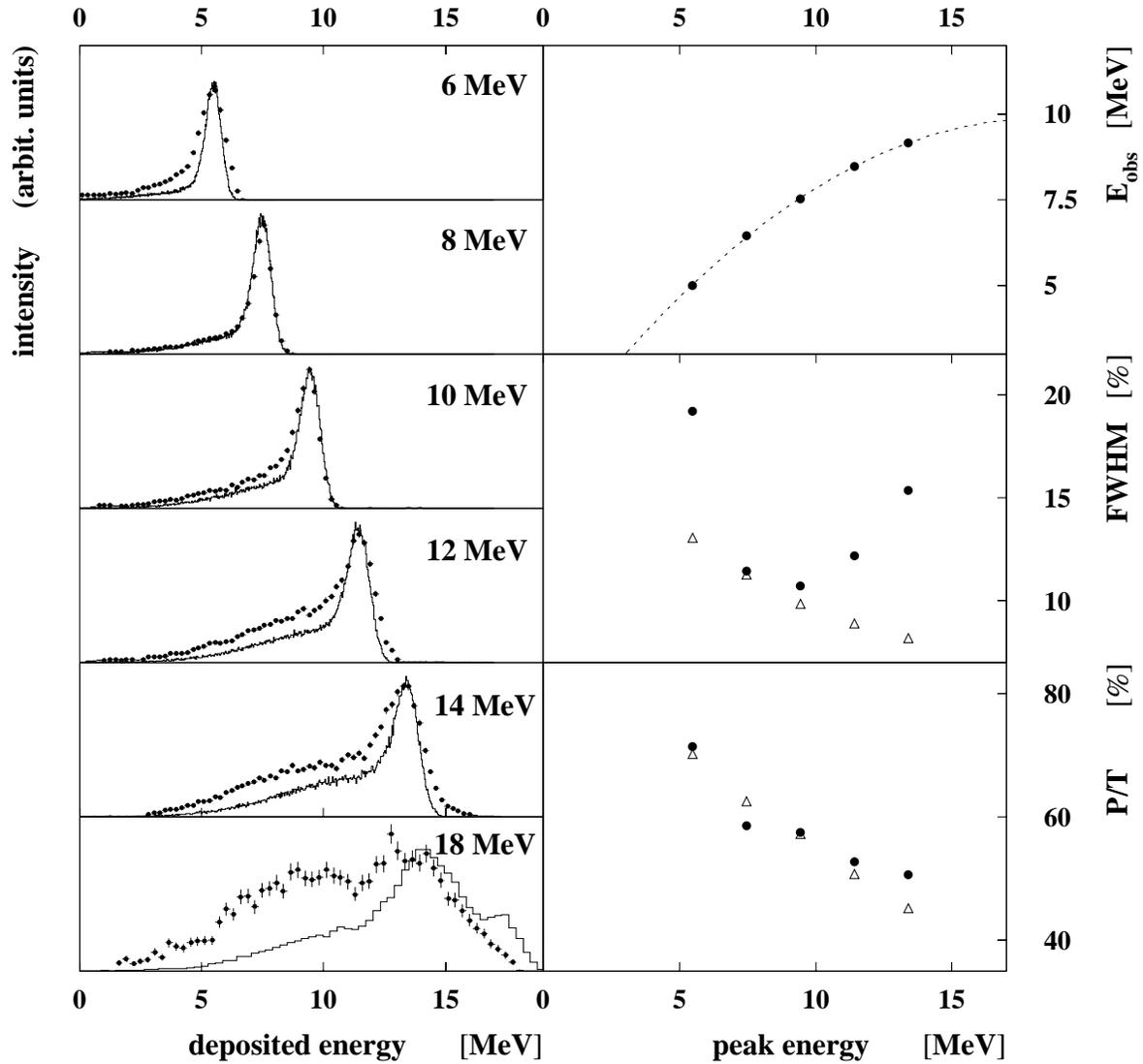,width=\textwidth}
\end{center}
\caption[]{ Measured detector response as a function of the deposited
electron energy. 
 Left: measured (data points) and simulated (histograms) line shapes for
 electron energies of 6, 8, 10, 12, 14, and 18 MeV. 
 Right: (top) dependence of the observed electron energy $E_{obs}$ 
 on the true energy, the dotted line represents the fit function 
 of Eq.\,\ref{eq:2_1-p41}, 
 (middle) the resolution (FWHM), and 
 (bottom) the peak to total ratio (P/T), as a function of the electron peak 
 energy in the experiment (dots) and in the GEANT simulations (triangles).}
\label{fig:linshap}
\end{figure}

The observed line shapes of all six small telescopes are identical within a few percent. Their energy signals show saturation
at higher incident energy as described by the equation: 

\begin{equation}       
E_{obs} = E_{dep} - \frac{E^{2}_{dep}}{k}. 
\label{eq:2_1-p41}
\end{equation}
Here E$_{dep}$ is  the deposited electron energy, E$_{obs}$ is the observed 
energy proportional to the ADC channel number and \mbox{$k$ is a} limiting 
energy parameter, which
was found to be about 40\,{MeV}.

From the GEANT simulation for 18 MeV electrons it is obvious
that only in a small fraction of the events the energy is fully absorbed. 
This is due to the limited depth (70 mm) of the E-detectors.  
The dominant peak around 14.5 MeV is due to energy loss.
Measurement and simulation show a satisfactory agreement for 
the measurements with 6 to 14 MeV electrons.
As the transition energy is shared among the two leptons, 
a full absorption efficiency up to 15 MeV for the individual
detectors is sufficient to investigate the desired range of
energies up to 18 MeV. 
When the range of the pair asymmetry $ y\, {\equiv}\, (E_1 - E_2)/(E_1 + E_2)$ 
can be limited to more symmetrical values (e.g. $|y| < 0.2$),
this range is extended to 25 MeV.

The above calibration implied the use of a focused electron beam
and normal impact on the detectors.
A comparison of the results thus obtained in Giessen with results from
the full experimental setup indicates that a larger number of electrons
are scattered out of the detectors, resulting in a decrease of the
peak to total ratio, from e.g.\ 0.6 at 8 MeV in the 
calibration run to 0.4 in the production run. Furthermore, slightly
different line shapes are expected for electrons and
positrons, especially in the low energy range \cite{meiring}.

\subsection{Solid angles and efficiencies of the detector telescopes}
\label{sec:deteffic}

The determination of the correlation angular distributions
required a good knowledge of the solid angles 
and the efficiencies of the telescopes. In view of this the telescopes 
were made equal to each other 
and were installed with carefully measured geometrical parameters (mean angle
and distance to target spot).

The solid angle acceptances of the telescopes were
calculated according to the geometrical parameters and dimensions of the
apparatus relative to the target position (see Table\,1).
The results were verified by simulations with the
GEANT code. From these simulations singles and coincidence
efficiencies were derived. In a next step, the combined
detector efficiencies and solid angles were verified by comparing
the measured singles telescope spectra 
with those obtained by the GEANT simulations. The multi-detector 
array, by its high redundancy, provides a comfortable way
to monitor the combined detector efficiencies by
permanently surveying the singles telescope spectra. 
According to this analysis, the total electron detection
efficiency of the detectors was larger than 90\% and equal (within a few
percent) for all telescopes.

The solid angles and the efficiencies of the later supplemented ``large"
telescopes were determined by fitting their correlation-angular
dependences relative to those extracted for the "small" telescopes.

\subsection{Results from the GEANT simulations}
\label{sec:Geantsim}

The three main sources for external pair conversion (EPC) in the experimental 
setup are the
target foil, the target frame and the carbon fibre tube. Since the EPC
cross-section is proportional to Z (Eq.\,5), care has 
been taken to use low Z components of minimized dimensions.
Due to the use of very thin targets, the EPC within the target
material represents a negligible fraction of all detected EPC events.
The contributions by the target frame are kept low by the tilt angle
of the telescopes relative to the target plane.
As expected, the
carbon-fibre tube, with its thickness of 0.8 mm is the dominant EPC source. 
This is verified by a projection of
the positions of pair conversion onto the instrumental components.
About 95\% of all EPC pairs, detected by a particular combination
of telescopes, are created
in a region inside the carbon fibre tube which has the size of a
$\Delta$E-detector. This region is located at mid-angle between the two 
telescopes. 
The result demonstrates the strongly forward
peaked angular distribution of EPC events as described in Section\,2.2.

In the case of the 17.23 MeV E1 transition in $^{12}$C we compared the 
theoretical IPC distribution with GEANT simulations including EPC and 
'out-scattering' and found good agreement at all angles (Section\,4.2). 

\subsection{Trigger for data readout and data-acquisition}
\label{sec:coeldaq}

The signals from the photomultipliers of the eight 
${\Delta}$E- and E-detectors were processed in constant fraction
discriminator units (CFD). The CFD-thresholds were adjusted
slightly above the noise level for the ${\Delta}$E-detectors, which are 
essentially insensitive to $\gamma$-ray events, and somewhat 
higher for the E-detectors. 
Chance events from double (or multiple) hits by $\gamma$-ray events 
in the
E-detectors are suppressed by about three orders of magnitude by requiring 
a ${\Delta}$E-E coincidence.

The resulting eight telescope signals are further processed 
in a CAMAC pattern-recognition module (MUEK), which has been developed 
at IKF\,\cite{muek}. 
In this module double telescope events are identified by their correlation
angle. 
Due to the small absolute solid angles
of the telescopes the rate of double telescope events
was much lower than the single telescope rates. In order to allow the
simultaneous measurement of single telescope events, the trigger module was
set to allow a 
scaled-down fraction of single telescope events. In this way 
all experiments could be run almost at the maximum beam currents available.
The spectra of single telescope events are used 
for on-line monitoring of the efficiencies and a 
first energy calibration of the E-detectors. 
Especially for the $\Delta$E-detectors with their 
low CFD thresholds this on-line survey was important.  
In the off-line 
analysis these spectra provide a reliable way to 
determine the telescope efficiencies, which are affected by 
the count-rate capability and stability 
of the electronics. 
Times and energies of all ${\Delta}$E- and E-detectors were recorded. 

\begin{figure}[htb] 
\leavevmode
\begin{center}
\epsfig{file=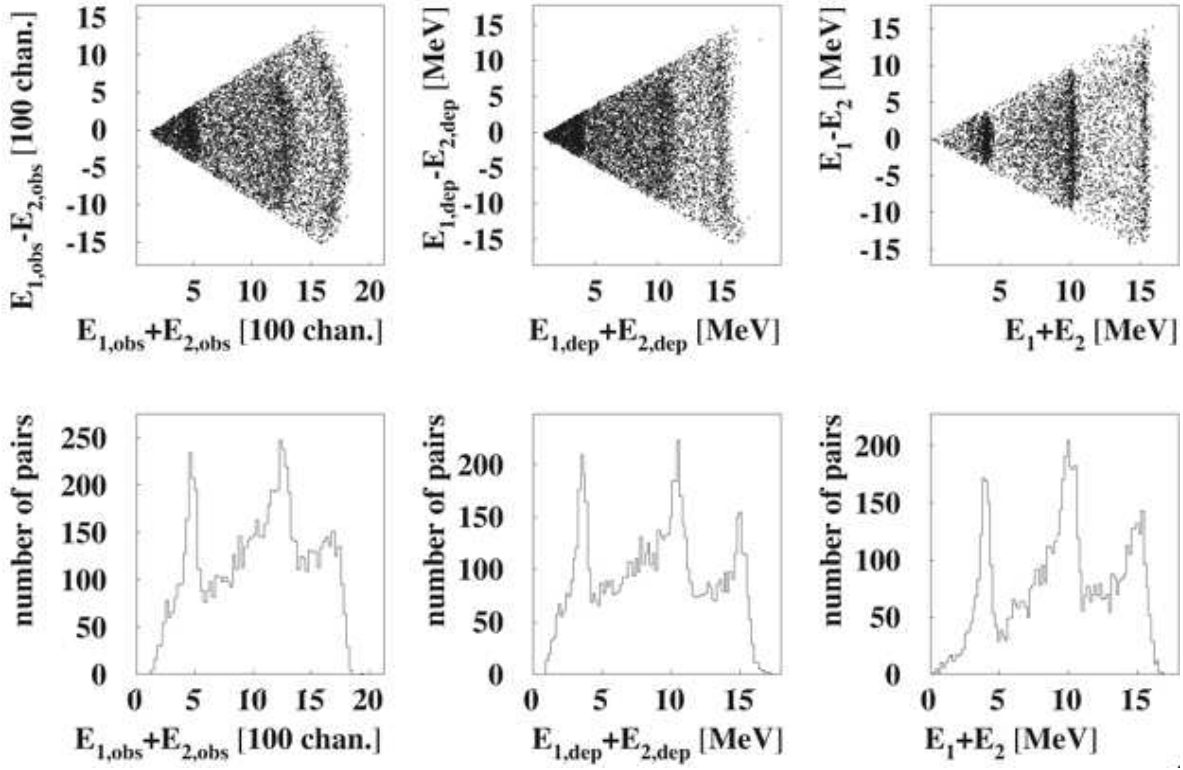,width=\textwidth}
\end{center}
\caption[]{ Two-dimensional representation of the sum-energy of detectors 1
and 2 against their energy difference.
 Top from left to right: observed energies in detector 1 and detector 2, the corresponding distribution of deposited  energies as determined from
Eq.\,7 and the result of a GEANT-simulation. Bottom: projection of the data 
in the upper row onto the sum energy axis. 
The GEANT spectra (right column) 
are folded with a Gaussian with FWHM = 0.28$\sqrt{\rm{E}}$ (E in MeV).} 
\label{fig:encorr}
\end{figure}

\section{Experimental Results}
\label{sec:expcond}

The spectrometer has been used in experiments at the 2.5-MV Van de Graaff 
accelerator of the IKF. Here we report on studies of the 
two well known proton induced reactions $^{11}$B$(p,e^+e^-)^{12}$C 
and $^{19}$F(p, ${\alpha} e^+e^-)^{16}$O at 1.6 MeV proton energy. 
For both reactions, angular correlations for pair emission can be found 
in the literature\,\cite{armbruster,devons1,devons2}. 
Both reactions could be observed simultaneously in our experiments, since a trace of fluorine 
contamination in the backing foil offered ample statistics for the 
E0 transition. The proton beam was focused to a beam spot size of 
approximately $3.0{\times}3.0$ mm$^{2}$ on the target.
Its intensity (between 10 and 17 $\mu$A) was limited by the count rate 
capability of the data processing, 
The cross-section of the $^{11}$B($p, e^+e^-)^{12}$C reaction shows
a wide resonance \mbox{(${\Gamma}\,{\approx}\,1$  MeV)} at a proton energy of
1.4 MeV with a maximum cross-section of about 27 ${\mu}$b/sr \cite{ajzenberger}.
At this energy two dominant transitions govern the $\gamma$-ray emission
spectrum \cite{huus}. One transition with E$_{\gamma}$ = 17.23 MeV depopulates the 17.23 (I$^{\pi} = 1^{-},$ T=1) MeV level to the $0^{+}$ ground state of 
$^{12}$C. The other one with E$_{\gamma}$ =  12.14 MeV depopulates
the 16.57 MeV (I$^{\pi} = 2^{-}$, T = 1) level to the first 
excited state\,\cite{armbruster}.
Both transitions are of E1 character, their relative contributions at the 
given incident energy of 1.6\,{MeV} and a target thickness of 600{\,}$\mu$g/cm$^{2}$ were estimated to yield $N_{17.23}/N_{12.14} \approx 2$. 
 
\begin{figure}[htb]
\leavevmode
\begin{center}
\epsfig{file=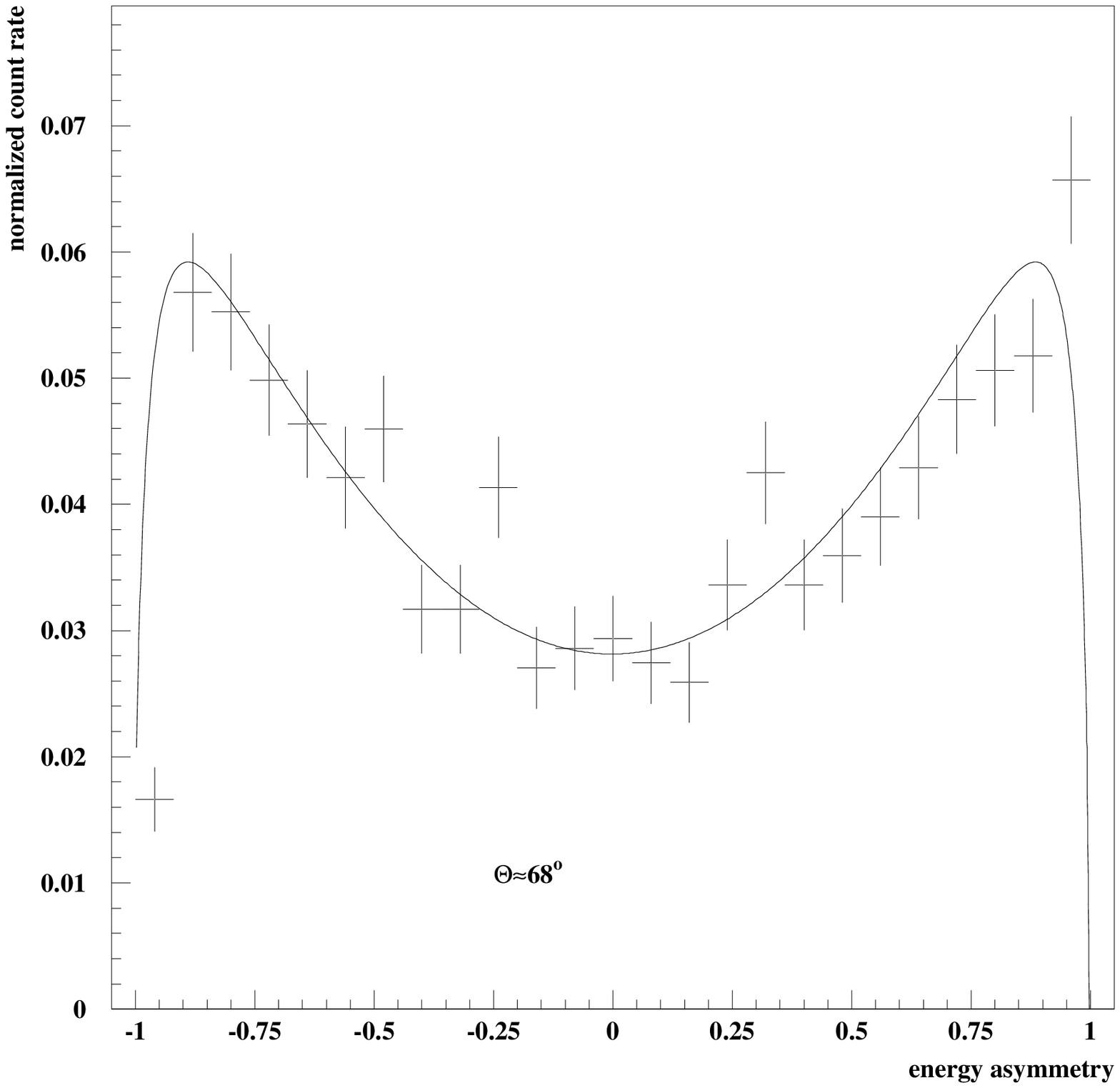,width=0.75\textwidth}
\end{center}
\caption[]{
 Energy asymmetry of pairs created by internal conversion in two M1 transitions at 14.64 and 17.64 MeV following the $^{7}$Li(p, e$^{+}$e$^{-}$)$^{8}$Be 
reaction at an opening angle of 68 degrees. The measured data (with vertical error bars)
have been normalized to the results from calculations in the
Born approximation (solid line).}
 \label{fig:asym}
\end{figure}

\subsection{Sum- and difference energy spectra}
\label{sec:analysis}

In the upper part of Fig.\,\ref{fig:encorr} correlation spectra 
for the detector combination 1 and 2 are shown for the
$^{11}$B(p,e$^+$e$^-$)$^{12}$C reaction at 1.6 MeV in their sum-energy 
versus difference-energy  representation. In the lower part of 
Fig.\,\ref{fig:encorr} the projection onto the sum-energy axis is given.
The structures at 11 and 15 MeV are due to the decay of the above 
mentioned two prominent excited levels in $^{12}$C{.} The deficit in energy 
for the 17-MeV transition (appearing at 15 MeV) is due to the  
energy signal saturation (see section 3.3).

\begin{figure}[t] 
\leavevmode
\begin{center}
\epsfig{file=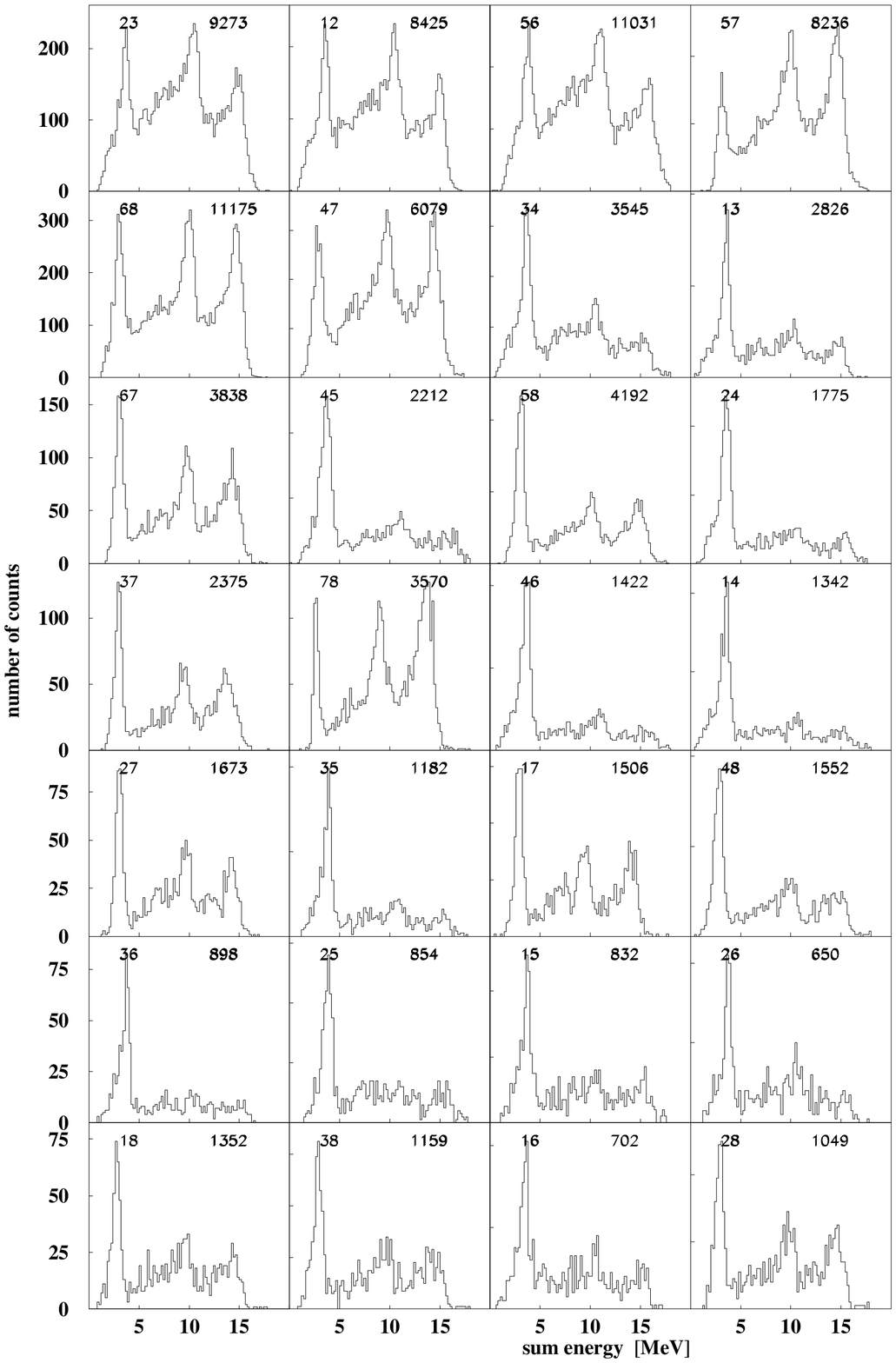,width=0.9\textwidth}
\end{center}
\caption[]{ Sum-energy spectra of electron-positron pairs from the $^{11}$B(p,e$^+e^-)^{12}$C reaction.
Detector combinations are ordered according to increasing correlation
angle. The total number of events is shown together with the corresponding 
combination number.} 
\label{fig:sumenergy}
\end{figure}

\begin{figure}[htb] 
\leavevmode
\begin{center}
\epsfig{file=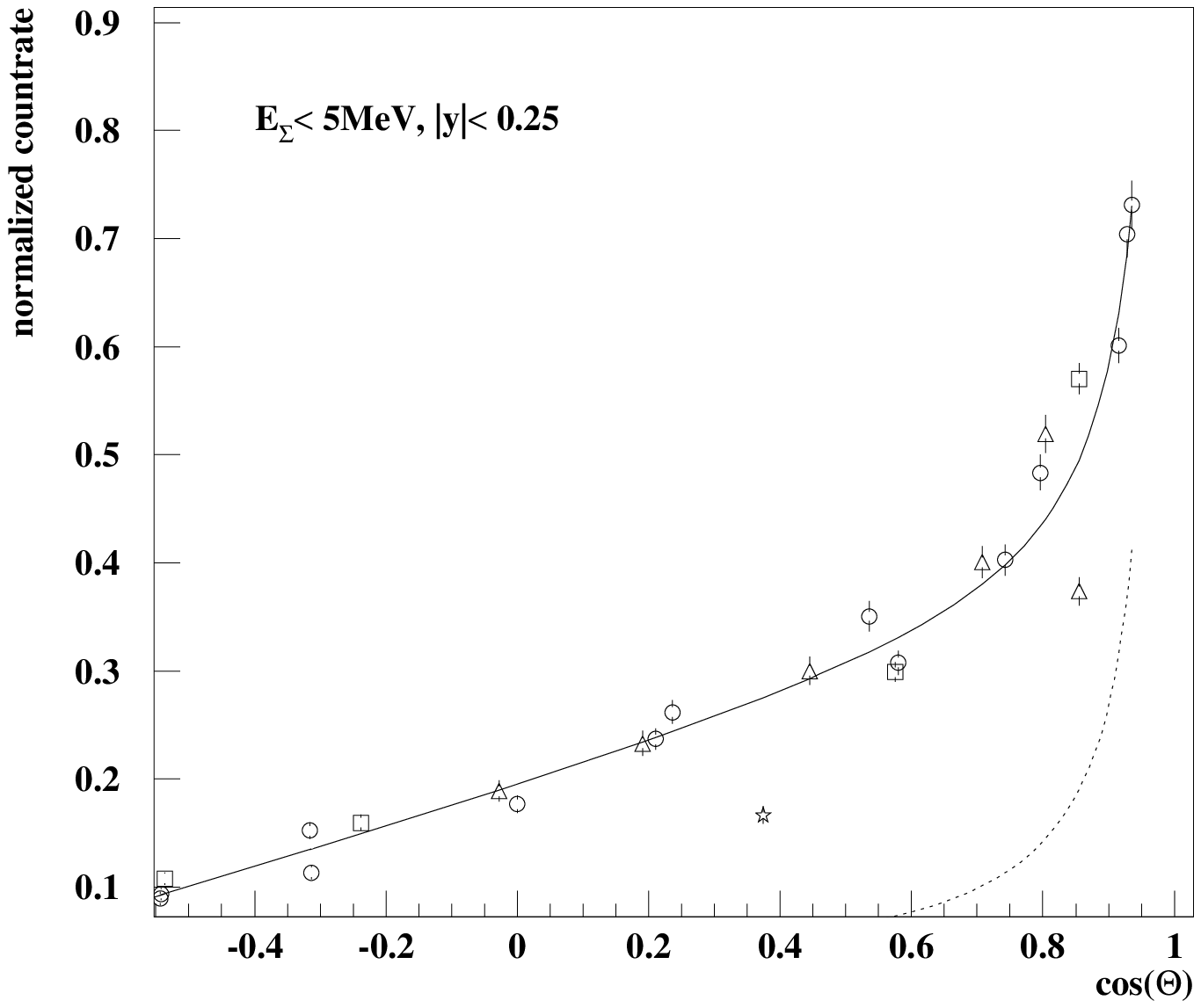,width=\textwidth}
\end{center}
\caption[]{Opening angle distribution of pairs with sum-energies
E$_\Sigma <$ 5 MeV from the reaction
$^{11}$B(p,e$^+$e$^-$)$^{12}$C. Solid circles represent 
Open circles are 
combinations between the small detectors, open
triangles between the small detectors and the large detector 7, open squares 
between the small detectors and the large detector 8, and the star between 
detectors 7 and 8. The solid line represents the
theoretical (1 + cos$\Theta$)-dependence for E0-transitions, normalized
 to the measured data.
In order to achieve good agreement at small correlation angles,
contributions from E1 transitions have been added (dotted line).
These contributions are due to low energy tails from the sum energy
lines at 11 and 15 MeV. } 
\end{figure}

The structure around 4 MeV is dominant over the whole angular
range and is ascribed to the 6.05\,MeV E0 transition 
in $^{16}$O. Its cross section results in 5\, mb at 1.6\, MeV. 
A contamination at the level of only 40 ng/cm$^2$ of
fluorine would explain the observed intensity. The
apparent energy deficit of approximately 1 MeV is explained by energy
loss in the carbon-fibre tube and is well reproduced by the  
GEANT simulations. 

In Fig.\,\ref{fig:asym} the energy difference of pairs is displayed
in terms of the energy \mbox{asymmetry $y$.} 
As the energy spectra at small correlation angles are contaminated by
EPC, we present here a spectrum from the combination of the two "large" 
telescopes which, in this experiment, had been mounted at a relative
angle of 68 degrees.  The spectrum is taken from a high-statistics run
of the $^7$Li(p,e$^{+}$e$^{-}$)$^8$Be reaction \cite{fokke0,fokke2,fokke3},
which follows a similar decay pattern as the $^{11}$B(p,e$^+$e$^-$)$^{12}$C
reaction and has two prominent M1 lines at 14.64 and 17.64 MeV. 
The spectrum is generated with a gate on the sum-energy peaks of the latter two
transitions. The theoretical expectations  are
well reproduced and demonstrate that the energy asymmetry spectrum is 
consistent with the conventional IPC process for M1.

\subsection{Distribution of opening angles}
\label{sec:openingangles}

\begin{figure}[htb]
\leavevmode
\begin{center}
\epsfig{file=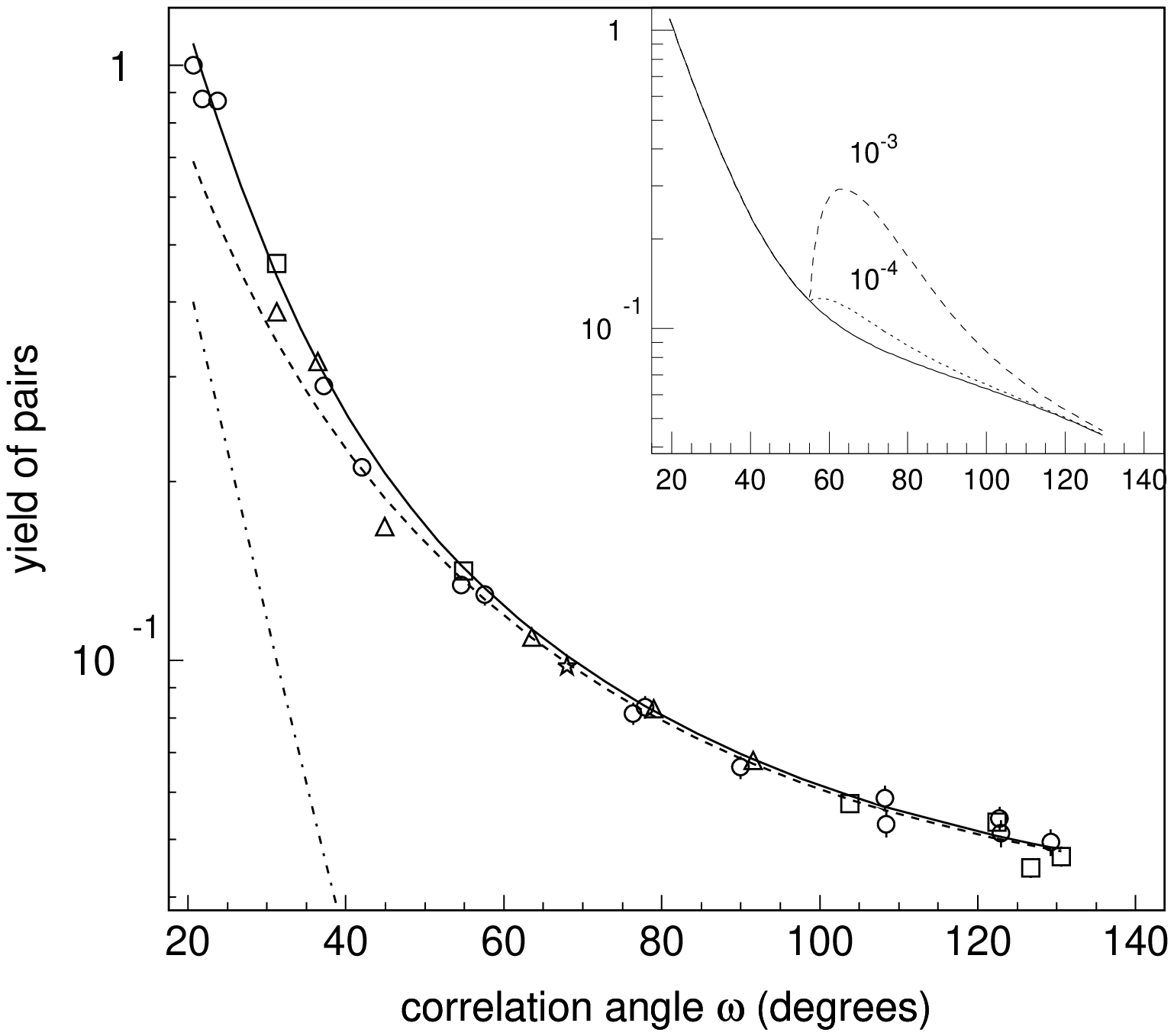, width=\textwidth}
\end{center}
\caption[]{Angular correlation (relative scale) of $e^{+}e^{-}$ pairs following the reaction 
$^{11}$B(p, $e^{+}e^{-})^{12}$C with sum energies larger than 5 MeV, 
using geometrical detector efficiencies. The theoretical 
E1-IPC correlation \cite{rose} (dashed line) has been normalized to the 
measured data in the angular range above 50$^{o}$. 
Open circles are 
combinations between the small detectors, open
triangles between the small detectors and the large detector 7, open squares 
between the small detectors and the large detector 8, and the star between 
detectors 7 and 8.  
The solid line includes effects from EPC \cite{olsen} and 
multiple scattering calculated in a Monte Carlo simulation (dot-dashed line).
In the insert the simulated shape is shown for the $e^+e^-$ decay
of a 9 MeV/c$^{2}$ boson with a branching ratio of 10$^{-4}$ and 
10$^{-3}$ relative to the 17.23 ${\gamma}$-ray emission, assuming isotropic 
emission of both the boson and the $e^{+}e^{-}$ pair.}  
\end{figure}

Fig.\,\ref{fig:sumenergy} shows the sum-energy spectra of the
boron measurement for the different telescope combinations. Since most of the 
leptons are within 
the minimum ionizing range, their energy loss in the target and the 
${\Delta}$E-detectors only results in a constant offset in the sum
energy spectra. 
The target thickness used in the
experiments was 600 ${\mu}$g/cm$^{2}$ and accounts for an energy
loss of 96 keV for 1.6 MeV protons{\,}\cite{janni}. 
This leads to different offsets for the three types of combinations;
small telescopes with small telescopes, small telescopes with large telescopes
and the combination of the two large telescopes. 
At closer inspection, the position of the 6.05 MeV structure varies slightly 
depending on the thickness of the two ${\Delta}$E-counters, as discussed in 
section 3.1.
Fig.\,6 also shows the large 
differences between the distributions of E0 and E1 transitions.
Whereas the
E0 contribution below 5 MeV is pronounced in all spectra,
the two E1 structures at sum-energies E$_{\Sigma}>$ 5 MeV
are strongest at small opening angles and are largely  
reduced at large correlation angles. 

The opening angle distribution for E0 transitions is expected to
follow a (1 + cos\,${\Theta})$-dependence. This is demonstrated
in  Fig.\,7 where the measured opening angle
distribution is plotted as a function of cos${\Theta}$. A good
correspondence to the theoretically expected dependence is
observed except for the (7-8) telescope combination which suffers from
a higher cut-off in the sum spectrum. 

Fig.\,8 shows the opening angle distribution for
the IPC contributions from the two E1-transitions populated in the
$^{11}$B(p,e$^+e^-)^{12}$C reaction. The data have been projected
with a threshold on the sum energy E$_{\Sigma}>$ 5 MeV. The $e^+e^-$
pair intensities have been scaled to unity at the smallest correlation 
angles.  
The dashed lines represent theoretical IPC distributions
\cite{rose} normalized to the data points at large $\Theta$
($>120^{o}$) where the contributions from EPC and multiple
scattering are minimal. The size of these latter contributions
(dot-dashed lines) has been estimated by means of the GEANT Monte
Carlo simulations using the EPC angular correlation from
ref.\,\cite{olsen} as generator for the $^{12}$C events
\cite{froehlich}. The solid line in Fig.\,8
represents the sum of all contributions.

In case of a hypothetical neutral boson with a mass of 9 MeV/c$^{2}$,   
we would observe a steep rise beginning at ${\Theta}=58^{o}$ and 
followed by a smooth decrease towards the larger angles. 
In the insert the signatures are displayed for the $e^{+}e^{-}$ decay 
of a 9 MeV/c$^{2}$ boson with branching ratio $B_X$ relative 
to ${\gamma}$-ray emission of $10^{-4}$ and $10^{-3}$. 
From the absence of any deviations from the conventional 
conversion processes in isovector E1 transitions, upper limits were derived 
on $B_X$ of scalar or vector bosons \cite{fokke0,fokke1,fokke2,fokke3}. 
They vary from $5.6{\cdot}10^{-5}$ to $1.0{\cdot}10^{-5}$ for a boson mass
between  6 to 15 MeV/c$^{2}$ with a value of $2.4{\cdot}10^{-5}$ for a 
9 MeV/c$^{2}$ boson at the 95\% confidence level. 
\section{Conclusion}
\label{conclusion}

A multi-detector array has been constructed to measure energy and
angular correlations of IPC from high-energy nuclear transitions. The 
information of all possible combinations of the eight lepton detector 
telescopes allows the simultaneous detection of $e^{+}e^{-}$ correlation angles between 20 and 130 degrees. The energy resolution of the lepton telescopes is
sufficient to identify and separate the nuclear transitions by
means of the sum energy and energy asymmetry for lepton pairs from 
transitions in the energy range between 5 and 18 MeV. By comparing the results
of various calibration experiments to extended Monte Carlo
simulations with the program code GEANT, it has been
shown that the spectrometer
is well suited to perform searches for an elusive short-lived neutral
boson with a mass between 6 and 15 MeV/c$^{2}$.

A series of dedicated experimental studies have already been 
performed with this setup. In one experiment, an excellent agreement  
with conventional IPC was found  
for the isovector E1 transitions at 12.2 and 17.2 MeV 
in $^{12}$C. These measurements result in an upper limit of $2.4{\cdot}10^{-5}$ 
for the branching ratio of a hypothetical boson with a mass of 9 MeV/c$^{2}$. 
The latter value is significantly below the theoretical pair conversion 
coefficient \cite{rose} of $3.9{\cdot}10^{-3}$ for a 17.2 MeV E1 transition, 
indicating the high sensitivity of the experimental device.

\section{Acknowledgements}

We acknowledge Th.\ W.\ Elze (IKF), K.\ Stelzer (IKF), R. van Dantzig (NIKHEF) and T.J. Ketel (NIKHEF) for valuable discussions. 
Many thanks are due to \mbox{K.\ Hildenbrand (GSI)} for his assistance
during the construction of the spectrometer, to M.\ Waldschmidt (IKF) for the 
excellently prepared targets and to \mbox{W.\ Arnold} and 
C.\ Wesselborg with their technical staff at the Giessen LINAC 
for help during the calibration runs. 
The members of the Nuclear Physics Group at ATOMKI, Debrecen, Hungary: 
\mbox{A. Krasznahorkay}, M. Csatl\'{o}s, Z. Gaczi, J. Guly\'{a}s, 
M. Hunyadi and Z. M\'{a}t\'{e} are acknowledged for the continuation
of the experimental program using the IKF-spectrometer at their MGC-20 cyclotron
\cite{atomki}.
One of the authors (K.\,A.\,M.) likes to thank the Deutsche 
Forschungsgemeinschaft
\, (DFG) for financial support in the frame of the Graduiertenkolleg Schwerionenphysik Giessen-Frankfurt. 


\end{document}